\documentclass[12pt]{article}

\usepackage{graphicx,epstopdf}
\usepackage{amsmath}
\usepackage{epsf}

\def\be{\begin{equation}}
\def\ee{\end{equation}}
\def\ba{\begin{eqnarray}}
\def\ea{\end{eqnarray}}
\def\ge{\mathrel{\raise.3ex\hbox{$>$\kern-.75em\lower1ex\hbox{$\sim$}}}}
\def\la{\mathrel{\raise.3ex\hbox{$<$\kern-.75em\lower1ex\hbox{$\sim$}}}}

\def\simgt{\mathrel{\raise.3ex\hbox{$>$\kern-.75em\lower1ex\hbox{$\sim$}}}}
\def\simlt{\mathrel{\raise.3ex\hbox{$<$\kern-.75em\lower1ex\hbox{$\sim$}}}}

\newcommand{\bi}[1]{\bibitem{#1}}
\newcommand{\fr}[2]{\frac{#1}{#2}}

\newcommand{\nc}{\newcommand}

\nc{\gone}{\bar g_{\pi NN}^{(1)}}
\nc{\gzero}{\bar g_{\pi NN}^{(0)}}
\nc{\al}{\alpha}
\nc{\ga}{\gamma}
\nc{\de}{\delta}
\nc{\ep}{\epsilon}
\nc{\ze}{\zeta}
\nc{\et}{\eta}
\nc{\ka}{\kappa}
\nc{\rh}{\rho}
\nc{\si}{\sigma}
\nc{\ta}{\tau}
\nc{\up}{\upsilon}
\nc{\ph}{\phi}
\nc{\ch}{\chi}
\nc{\ps}{\psi}
\nc{\om}{\omega}
\nc{\Ga}{\Gamma}
\nc{\De}{\Delta}
\nc{\La}{\Lambda}
\nc{\Si}{\Sigma}
\nc{\Up}{\Upsilon}
\nc{\Ph}{\Phi}
\nc{\Ps}{\Psi}
\nc{\Om}{\Omega}
\nc{\ptl}{\partial}
\nc{\del}{\nabla}
\nc{\ov}{\overline}
\nc{\newcaption}[1]{\centerline{\parbox{15cm}{\caption{#1}}}}
\nc{\uop}{U(1)$'$}

\def\beq{\begin{equation}}
\def\eeq{\end{equation}}
\def\bmat{\begin{displaymath}}
\def\emat{\end{displaymath}}
\def\bear{\begin{eqnarray}}
\def\eear{\end{eqnarray}}
\def\ba{\begin{eqnarray}}
\def\ea{\end{eqnarray}}
\def\bery{\begin{array}}
\def\ery{\end{array}}
\def\bit{\begin{itemize}}
\def\eit{\end{itemize}}
\def\ben{\begin{enumerate}}
\def\een{\end{enumerate}}
\def\btab{\begin{tabular}}
\def\etab{\end{tabular}}
\def\btbl{\begin{table}}
\def\etbl{\end{table}}
\def\bfig{\begin{figure}[htb]}
\def\efig{\end{figure}}
\def\bpic{\begin{picture}}
\def\epic{\end{picture}}

\def\ga{\mathrel{\raise.3ex\hbox{$>$\kern-.75em\lower1ex\hbox{$\sim$}}}}
\def\la{\mathrel{\raise.3ex\hbox{$<$\kern-.75em\lower1ex\hbox{$\sim$}}}}
\def\gappeq{\mathrel{\rlap {\raise.5ex\hbox{$>$}}
{\lower.5ex\hbox{$\sim$}}}}
\def\lappeq{\mathrel{\rlap{\raise.5ex\hbox{$<$}}
{\lower.5ex\hbox{$\sim$}}}}

\def\gyr{{\rm \, G\kern-0.125em yr}}
\def\mev{{\rm \, Me\kern-0.125em V}}
\def\gev{{\rm \, Ge\kern-0.125em V}}
\def\tev{{\rm \, Te\kern-0.125em V}}

\def\slash#1{\rlap{\hbox{$\mskip 1 mu /$}}#1}%

\begin{document}

\begin{titlepage}

\setcounter{page}{1}

\vspace*{0.2in}

\begin{center}

\hspace*{-0.6cm}\parbox{17.5cm}{\Large \bf \begin{center}
Secluded U(1) below the weak scale\\
\end{center}}

\vspace*{0.5cm}
\normalsize

{\bf  Maxim Pospelov }

\smallskip
\medskip

$^{\,(a)}${\it Perimeter Institute for Theoretical Physics, Waterloo,
ON, N2J 2W9, Canada}

$^{\,(b)}${\it Department of Physics and Astronomy, University of Victoria, \\
     Victoria, BC, V8P 1A1 Canada}

\smallskip
\end{center}
\vskip0.2in

\centerline{\large\bf Abstract}

A secluded U(1) sector with weak admixture to photons, $O(10^{-2}-10^{-3})$, 
and the scale of the breaking below 1 GeV represents a natural yet poorly 
constrained extension of the Standard Model.  We analyze $g-2$ of muons and 
electrons together with other precision QED data, as well as radiative decays of 
strange particles to constrain mass--mixing angle ($m_V-\kappa$) parameter space. 
We point out  that $m_V \simeq 214$ MeV 
and $\kappa^2 > 3\times 10^{-5}$ can be consistent with the hypothesis of HyperCP collaboration,
that seeks to explain the anomalous energy distribution of muon pairs in the $\Sigma^+ \to p \mu^+\mu^-$ process
by a resonance, without direct contradiction to the existing data on radiative kaon decays. The same parameters
lead to $O( {\rm few} \times 10^{-9})$ upward correction to the 
anomalous magnetic moment of the muon, possibly relaxing some tension between experimental 
value and theoretical determinations of $g-2$. The ultra-fine energy resolution scan of $e^+e^-\to \mu^+\mu^-$
cross section and dedicated analysis of lepton spectra from $K^+\to \pi^+ e^+e^-$ 
decays should be able to provide a conclusive test of this hypothesis and improve the constraints on the model. 

\vfil
\leftline{November 2008}

\end{titlepage}

\section{Introduction}

Extra U(1)$'$ group(s) represent a rather minimal and in some sense natural 
extension of the Standard Model. The most economical way of making this sector 
"noticeable" to the Standard Model (SM) particles and fields is via the so-called 
kinetic mixing portal, which is simply a coupling between 
the U(1) of the SM hypercharge and U(1)$'$ \cite{Holdom}:
\be
{\cal L}_{ \rm SM+U(1)'} = {\cal L}_{ \rm SM}-
\fr{1}{4}V_{\mu\nu}^2 + 
\fr12\kappa V_{\mu\nu} F^Y_{\mu\nu}
+{\cal L}_{\rm Higgs'}+...
\label{VdF}
\ee
Here $V_{\mu\nu}$, $F^Y_{\mu\nu}$ are the field strengths of the U(1)$'$ and U(1)$_{\rm SM}$, 
and $\kappa$ is the mixing angle. The specific form of 
the U(1)$'$ Higgs sector is not so important, but for simplicity we shall assume that the 
breaking occurs due to some elementary Higgs$'$ field with the Mexican hat potential.
The ellipses stands for other possible matter fields, singlet under the SM 
and charged under the \uop. In particular, the ellipses may include the weakly interacting massive 
particles (WIMPs) charged under \uop, in which case $V-\gamma$ mixing mediates interaction 
between visible and dark matter sectors \cite{PRV1}.

Such coupling, and an overall neutrality of the SM under U(1)$'$ ensure the absence of 
problems with anomalies. This is, of course, not the only possibility of introducing
$Z'$ physics, and other examples with and without supersymmetry have been proposed and 
studied at length (See, {\em e.g.} \cite{Fayet1,Langacker} and Refs. therein). 
If the mass of the exra U(1)$'$ gauge boson is at the TeV scale, only the sizable coupling to the
SM would allow for the collider tests of such models. It is quite natural, however,
to consider the following range of parameters that allows to seclude the \uop\ sector and place it well 
below the electroweak scale:
\begin{eqnarray}
\label{relation}
\alpha' \sim \alpha_{\rm SM}(M_Z); ~~~ \kappa \sim (\alpha \alpha')^{1/2}/\pi; ~~~ m_V^2 \sim {\rm loop}
\times\kappa^2 M_Z^2.
\end{eqnarray}
Such values of $\kappa \sim O(10^{-2}-10^{-3})$ can be induced radiatively
by the loops of unspecified very heavy 
particles charged under both U(1) groups. This way the mixing parameter will
depend on the logarithm of the ratio of some UV scale ({\em e.g.} GUT scale)
to the mass scale of particles charged under both groups. 
The last relation in (\ref{relation}) 
implies a "radiative transfer" of the gauge symmetry breaking from the SM to the \uop. 
For our range of $\kappa$, it suggests that the mass scale of the vector particle is under 1 GeV. 
This line of arguments justifies a closer look at the phenomenology of low-scale \uop\ models with the 
kinetic mixing to photons. 

In recent years, some interest to the $m_V\sim O($MeV-GeV), per-mill coupling gauge boson physics 
has been driven by the model-building attempts to construct WIMP models with masses in the 
MeV range  \cite{PRV1,BF,Kim}. Some particle phenomenology aspects of the mediator physics  
have been discussed in Refs. \cite{PRV1,BouchF}, and most notably by Fayet in \cite{Fayet2}. 
Most recently, an independent motivation for light mediators for the 
TeV-scale WIMPs have been advocated in \cite{recent_stuff} as the most natural way of having an 
enhanced annihilation in Galactic environment. This speculation is fueled by recent results of 
PAMELA collaboration \cite{Pamela}, that sees evidence for an enhanced fraction of 
high-energy positrons that may have been created through WIMPs. For other 
investigations of the vector model with kinetic mixing, covering different 
phenomenological aspects and different parameter range,  see {\em e.g.} recent works \cite{lowV}. 

Leaving the WIMP physics aside, the purpose of this note is to investigate the phenomenology 
of MeV-to-GeV scale mediators, keeping both $\kappa$ and $m_V$ as free parameters. In Sections 2 and 3
we will address the constraints coming from the anomalous magnetic moments of electron and muon,
as well as other precision QED tests, and the signatures of secluded \uop\ in the decays of strange particle. 
We reach our conclusions in Section 4.

\section{QED tests of secluded U(1)}

Since we are going to investigate the MeV-scale phenomenology of $V$-bosons, 
only their mixing with photons is relevant. Retaining the photon part of $F^Y_{\mu\nu}$, redefining 
$\kappa$ to absorb the dependence on $\theta_W$, assuming the 
breaking of \uop, and using the equations of motion, we arrive
at the following effective Lagrangian,
\begin{eqnarray}
{\cal L}_{ \rm eff} = -\fr{1}{4}V_{\mu\nu}^2 +\fr{1}{2} m_V^2 V_\mu^2 + \kappa V_\nu \partial_\mu F_{\mu\nu}
+{\cal L}_{h'}+... \nonumber\\\!\!\!
= -\fr{1}{4}V_{\mu\nu}^2 +\fr{1}{2} m_V^2 V_\mu^2 + \kappa e  J_\mu V_\mu 
+{\cal L}_{h'}+...,
\label{eff}
\end{eqnarray}
where in the second line 
the divergence of the photon field strength is traded for the operator of
the electromagnetic current. The ${\cal L}_{h'}$ term represents the Lagrangian of the Higgs$'$ particle. 
As evident from (\ref{eff}),  the production or decay of $V$-bosons occur via the intermediate 
"non-propagating" photon, as the $q^2$ in the $\kappa$-insertion cancels $1/q^2$ of the photon propagator. 
A simple examination of Lagrangian (\ref{eff}) reveals two distinct dynamical regimes for the processes 
mediated by the exchange of virtual $V$. When the $q^2$ of momentum flowing through the $V$ line is much larger
than $m_V^2$, the $V$-exchange is analogous to the photon exchange, and thus leads to a 
simple renormalization of the fine structure constant. For the momenta much smaller than 
$m_V$, the exchange of $V$-boson introduces an additional current-current contact interaction,
that mimics the contribution of particle's charge radius \cite{PRV1}. 
\begin{figure}[htbp]
\centerline{\includegraphics[width=9.8cm]{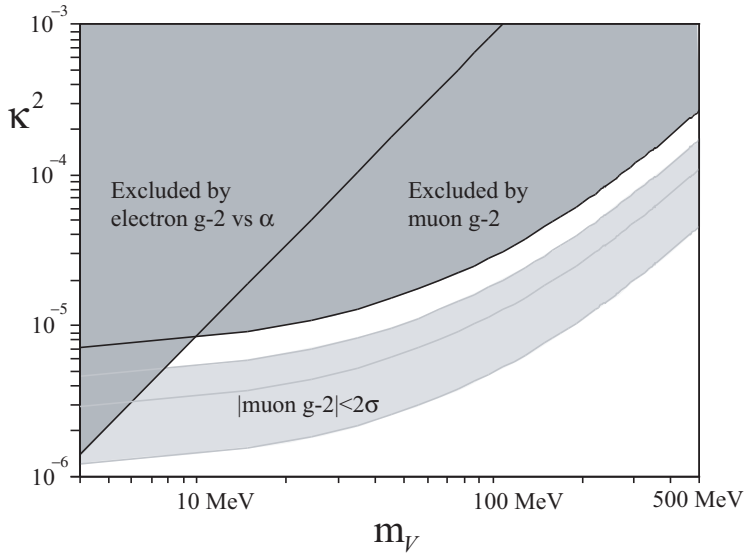}}
 \caption{\footnotesize  
Combination of $g-2$ and $\alpha$ measurement constraints on $m_V-\kappa^2$ parameter space. 
The dark grey color indicate the excluded region. The light grey band is where the consistency 
of theoretical and experimental values of $(g-2)_\mu$ improves to $2\sigma$ or less. The grey line inside this band
indicate $0\sigma$ relative to experimental value, {\i.e.} a positive shift of $3\times 10^{-9}$ to $a_\mu^{\rm th}$.
}
\label{f1} 
\end{figure}

Calculation of the one-loop diagram produces the result for 
the additional contribution of $V$ bosons to the 
anomalous magnetic moment of a lepton (electron, muon) $a_l^{V}$, that can be conveniently 
expressed as:
\be
\label{g-2}
a_l^{V}  = \fr{\alpha}{2\pi} \times \kappa^2 \int_{0}^{1}dz\fr{2m_l^2z(1-z)^2}{m_l^2(1-z)^2+m_V^2z}
=\fr{\alpha\kappa^2}{2\pi}\times \left\{\begin{array}{c}1 ~~{\rm for} ~~ m_l\gg m_V,\\
										2m_l^2/(3m_V^2) ~~{\rm for} ~~ m_l\ll m_V.\end{array}
\right. 
\ee
We introduce the notation $F(m_l^2/m_V^2)$ for the integral in (\ref{g-2}).

Currently, the precision measurement of $(g-2)_e$ \cite{Harvard} surpasses the sensitivity of all other QED
measurements, and is used for the extraction of the fine structure constant \cite{GK}. Therefore, 
Eq. (\ref{g-2}) can be re-interpreted as an effective shift of the coupling constant by 
\be
\label{dalpha}
\Delta \alpha = 2\pi a_e^V; ~~ \Delta \alpha^{-1} = - 2\pi a_e^V/\alpha^2,
\ee
and the precision test of the model comes from the {\em next} most precise 
determination of $\alpha$. Currently, these are atomic physics results with Cs and Rb 
\cite{CsRb}. These determinations are very weekly affected by the additional $V$ boson,
compared to $(g-2)_e$. Adopting the results of \cite{CsRb}, we require that the relative 
shift of $\Delta \alpha$ does not exceed 15 ppb, which results in the following constraints 
on the parameters of our model:
\be
\label{ge}
\kappa^2 \times F(m_e^2/m_V^2)  < 15\times 10^{-9} ~~ \Longrightarrow ~~ 
\kappa^2\times \left( \fr{\rm 100~MeV}{m_V}  \right)^2 < 1. \times 10^{-3},
\ee
where we also made a relatively safe assumption that $m_V \gg m_e$. In practice one has to require
$m_V \ga 4$ MeV in order to satisfy  constraints imposed by primordial nucleosynthesis (BBN) \cite{RS}. 
If $m_V$ is chosen right at the boundary of the BBN constraint, 
Eq. (\ref{ge}) requires $\kappa^2 $ to be less than $10^{-6}$, 
while of course the constraint weakens considerably for larger values of $m_V$.

Another important constraint comes from the measurement of the 
muon magnetic anomaly. The application of this constraint is not straightforward
due to the necessity to deal with hadronic uncertainty in extracting theoretical prediction for 
$a_\mu$. The determination based on $e^+e^-$ annihilation to hadrons 
points to a $+302(88)\times 10^{-11}$ deficit 
(see, {\em e.g.} \cite{PMS} and references therein)
of $a_\mu^{\rm th}$ relative to the experimental value for $a_\mu^{\rm exp}$ measured at Brookhaven \cite{gmu}. 
This constitutes a $3.4\sigma$ deviation, which over the years 
has prompted numerous theoretical speculations 
on new physics "solution" to this discrepancy. 
Other determinations based on $\tau$ physics \cite{tau}, and most recently on 
preliminary analysis of the radiative return at BaBar \cite{DavierNsk} 
do not indicate any discrepancy. It is easy to see that the 
positivity of $V$-contribution (\ref{g-2}), improves the agreement between $a_\mu^{\rm th}$
and $a_\mu^{\rm exp}$, if one adopts the $e^+e^-\to$hadrons based result. 
To state a conservative limit, we require that 
$a_\mu ^V \le (302+5\sigma)\times 10^{-11}
= 7.4\times 10^{-9}$. Such additional contribution to the anomaly is excluded no matter what 
method of treating the hadronic contribution to $a_\mu^{\rm th}$ one would like to choose. 
The combination of $g-2$ constraints is shown in Figure 1. The muon constraints include 
a forbidden region, as well as a "welcome" band of $  1.3\times 10^{-9}< a_\mu^V < 4.8\times 10^{-9}$ that puts the theoretical 
prediction based on $e^+e^-\to$hadrons within two standard deviations from the experimental result $a_\mu^{\rm exp}$.

Another possibility for probing $V-\gamma$ mixing is through the high-precision measurements of 
photon exchange. For example,  the $V$-contribution to the electron-proton scattering amplitude at 
$|q^2| \ll m_V^2$ is equivalent to the $(r_c^V)^2 = 6\kappa^2/m_V^2$ correction to the proton charge radius. 
The high-precision measurements of the Lamb shift are used to extract the proton charge radius
(see {\em e.g.} Ref. \cite{Sav} and references therein), 
but in order to test $V$-exchange induced contribution, one has to measure the same quantity, 
$r_c^2$ using different techniques in the kinematic regime 
where $|q^2|$ would be on the order or larger than $m_V^2$. 
Since  the measurements of charge radii in scattering are intrinsically less precise 
than the Lamb shift determination of $r_c$, the constraint from the charge 
radius of the proton does not appear to be better than 
\be
\fr{6\kappa^2}{m_V^2} \la 0.1 {\rm fm}^2~~ ~~ \Longrightarrow ~~ 
\kappa^2\times \left( \fr{\rm 100~MeV}{m_V}  \right)^2 < 4\times 10^{-3},
\ee
and as such is subdominant to the $(g-2)_e-\alpha$ constraint (\ref{ge}). It remains to be 
seen whether other precision QED tests ({\em e.g.} involving muonic atoms) would be able to 
improve on this constraint.

Finally, the  $V-\gamma$ mixing may be searched for as a  narrow resonance in the 
$e^+e^-$ collisions. The quantum numbers of $V$ allow them to be seen as narrow sharp resonances 
in the $s$-channel. The leptonic widths of this resonance can be easily calculated to be  
\be
\label{Gammal}
\Gamma_{e^+e^-} = \fr13 \kappa^2 \alpha m_V,~~
\Gamma_{\mu^+\mu^-} = \fr13 \kappa^2 \alpha m_V\left(1+\fr{2m_\mu^2}{m_V^2}\right)\sqrt{1-\fr{4m_\mu^2}{m_V^2}}.
\ee
Very close to the muon threshold the second formula will be modified by the Coulomb interaction of the outgoing
muons. Above the hadronic threshold, $\Gamma_{\rm hadr}$ is 
directly related to the total rate of $e^+e^-$ annihilation 
into hadrons at the center of mass energy equal to $m_V$.
For any value of parameters $V$-resonance is extremely narrow, 
$\Gamma_{e^+e^-} = 2.4~{\rm keV}\times \kappa^2 (m_V/100~{\rm MeV})$,
which is smaller than the typical energy spread for the colliding particles.
Near the resonance, however, the cross sections for {\em e.g.} $e^+e^-\to e^+e^-$ 
or $e^+e^-\to \mu^+\mu^-$  may be significantly enhanced relative to the  
usual QED cross section for $e^+e^-\to \mu^+\mu^-$. Therefore, a proper procedure would be to 
compare resonant and standard non-resonant cross section smeared over some typical 
energy interval $\Delta E$, provided of course that the resonant energy is
within this interval $\Delta E$.  For example, for $e^+e^-\to \mu^+\mu^-$ process 
we have 
\be
\label{ratio}
\fr{\sigma^{\rm res}_{e^+e^-\to \mu^+\mu^-}}{\sigma^{\rm standard}_{e^+e^-\to \mu^+\mu^-}} \simeq 
\frac{9\pi\Gamma_{\rm tot}}{8\alpha^2 \Delta E} \times 
\fr{{\rm Br}_{e^+e^-}{\rm Br}_{\mu^+\mu^-}}{(1+2m_\mu^2/m_V^2)\sqrt{1-4m_\mu^2/m_V^2}},
\ee
where ${\rm Br}_{l^+l^-}$ are the leptonic branching ratios of $V$-resonance, and $\Gamma_{\rm tot}$ its
total width. 
These branching ratios are comparable to 1 and the total width is given by the sum of
the leptonic and hadronic widths {\em if} there are no matter fields in the 
\uop -charged sector with masses less than $m_V/2$. In this case, 
ratio (\ref{ratio}) scales as $\kappa^2 m_V/(\alpha\Delta E)$, which can provide an important constraint 
on the model. Choosing somewhat conservatively $\Delta E \sim$ 1 MeV, and 
requiring this ratio to be less than 0.1, we plot this sensitivity level in Figure 2, 
which also includes other conditional 
constraints from the next section. With these assumptions, the $e^+e^-$  
scattering can become the most precise probe for $m_V > 100$ MeV, and could rule out $O(10^{-3})$ level 
of mixing. 
At this point we will refrain from calling it a constraint, as we believe that in practice, without 
a dedicated search, $O({\rm meV-eV})$ width resonances can be missed even if they are very strong. 
It is also important to  keep in mind that if there exist 
additional channels for $V$ to decay into the matter charged under \uop, the ratio (\ref{ratio})
scales as $\sim \kappa^4$, and all constraints weaken considerably. This is exactly the 
case in models with MeV-scale dark matter, where decays to two dark matter 
particles can make $\Gamma_V$  parametrically larger than (\ref{Gammal}) \cite{Fayet2}.

\section{Production of \uop\ bosons in hyperon and $K$ decays}

Radiative decays of strange particles is another natural place 
where \uop\ bosons can manifest themselves. Since the interaction 
is mediated by mixing with the photon, the natural place to look 
for $V$ are the decays of $K$ and $\Sigma^+$ with photon or lepton pair(s) in the final state. A lot of work has been 
done in this area over the years, and the most important conclusion 
is that the rates for the flavor-changing radiative decays are 
dominated by large-distance physics, which cannot be (or almost cannot be) extracted from first principles. 
For this paper we employ the following strategy: we use the existing evaluations 
of the kaon and hyperon vertices with on-shell or off-shell photons, and use them to calculate 
the production of $V$-bosons. We also notice that the vector particles in our model have interactions
only with conserved current, which makes them extremely difficult to produce in flavor-changing transitions
as opposed to mediators with {\em e.g.} scalar \cite{PRV1,Bird} or axial-vector couplings \cite{Fayet2}. 

\subsection{ Radiative Kaon decays } 

In this paper we will consider two important processes,
\begin{eqnarray}
\label{Kproc}
{\rm A:}~~ K^+ &\rightarrow& \pi^+ V~~~[K^+ \rightarrow \pi^+ l^+ l^-]\\
{\rm B:}~~ K^+&\rightarrow& l^+\nu V~~~[K^+ \rightarrow l^+\nu,~~K^+ \rightarrow l^+\nu l^+ l^-]\nonumber,
\end{eqnarray}
where the SM processes are shown inside square brackets. The branchings for the SM 
processes with $l^+ l^-$ are small, on the order of $O(10^{-7}-10^{-8})$ depending on 
particular process of interest. For the semileptonic decays A, 
the SM rates were estimated in Ref. \cite{dAmbrosio}, 
where the starting point was the chiral perturbation theory together with the
experimental input for the $K^+\to \pi^+\pi^-\pi^+$ vertex. This analysis results in the 
prediction for the $q^2$-proportional vertex of $K-\pi$ transition with virtual 
photon. In terms of this vertex, in the notation of Ref. \cite{dAmbrosio},
the expression for the amplitude is 
\be
{\cal M}_{K\to\pi V} = \fr{e\kappa m_V^2}{(4 \pi)^2m_K^2}(k+p)_\mu\epsilon^V_\mu W(m_V^2),
\ee
where $k$ and $p$ are the kaon and pion momenta, $\epsilon^V_\mu$ is the polarization 
of $V$-boson, and $W^2(m_V^2) \simeq 10^{-12} (3+ 6m_V^2/m_K^2)$ \cite{dAmbrosio}. 
The latter is in reasonable agreement with experimental determination 
via the $K^+\to\pi^+e^+e^-$ decay \cite{Appel} and with the rate of $K^+\to\pi^+\mu^+\mu^-$ decay \cite{Park}.
Notice the proportionality of the amplitude to $m_V^2$ that replaces $q^2$ of the virtual photon 
and suppresses the rate for small $m_V\ll m_K$. 
This amplitude gives rise to the following branching ratio:
\be
\Gamma_{K\to\pi V} = \fr{\alpha\kappa^2}{2^{10}\pi^4} \fr{m_V^2W^2}{m_K}f(m_V,m_K,m_\pi)
~~\Longrightarrow~~{\rm Br}_{K\to\pi V} 
\simeq 8\times 10^{-5}\times \kappa^2 \left( \fr{m_V}{\rm 100~MeV}  \right)^2.
\label{Ktopi}
\ee
In this formula, dimensionless factor 
$f(m_V,m_K,m_\pi)$ stands for the mass dependence of phase space and matrix element, and
$f$ is normalized to 1 in the limit $m_{\pi,V} \to 0$ when $m_K$ is kept finite. 
The last relation in (\ref{Ktopi}) is valid only when $m_V$ is much smaller than $m_K$, but in practice for 
all $m_V$ below 200 MeV. 

 \begin{figure}[htbp]
\centerline{\includegraphics[width=10cm]{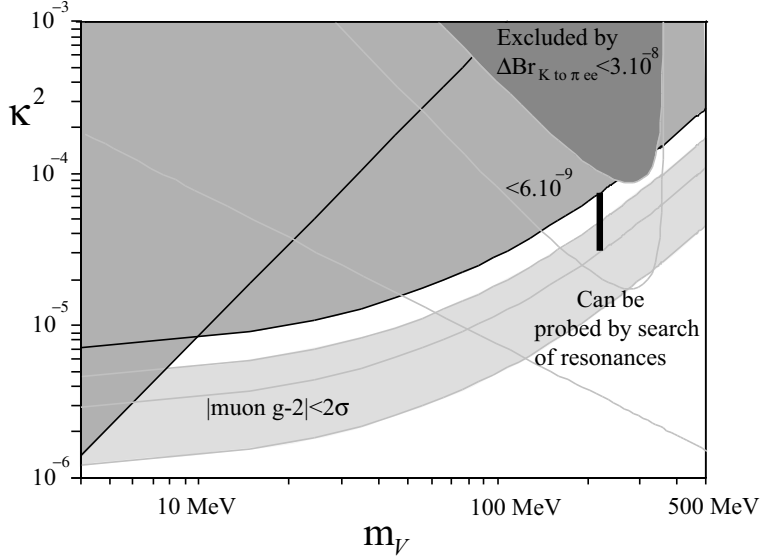}}
\vspace*{0.1in}
 \caption{\footnotesize  Same as Figure 1, but with some conditional constraints in the assumption 
of purely "visible" decays of $V$. The darkest grey region is from 
${\rm Br}^V_{K^+\to \pi^+e^+e^-} < 3.\times 10^{-8}$; and the similarly shaped grey line is possible to achieve 
with re-analysis of $V_{K^+\to \pi^+e^+e^-}$ at $\Delta$Br$<6\times 10^{-9}$ level. The grey diagonal straight line 
indicates the level of sensitivity that can be achieved via the $e^+e^-$ search of extremely narrow resonances. The thick
vertical bar indicates the region consistent with the HyperCP hypothesis (\ref{whatyouneed})
and other constraints.}
\label{f2} 
\end{figure}

In order to constrain (\ref{Ktopi}), one has to know the subsequent fate of $V$. It can decay 
to lepton pairs or invisibly, if such channel is open. In case of the invisible decay, 
one could use the results of $K^+\to\pi^+\nu\bar\nu$ search, but due to 
a rather restrictive kinematic window for pion momentum \cite{invisible},
this constraint is difficult to implement for arbitrary $m_V$. If the invisible decay is 
absent, $K^+\to\pi^+ V\to \pi l^+ l^-$ decays will contribute to the $K^+ \rightarrow \pi^+ l^+ l^-$
process. Given that there is still some uncertainty in the determination 
of $W(q^2)$ and its shape, and without a dedicated search for a resonant part, 
one could still contemplate that $\sim$10\% of the existing branching ratio
for $K^+\to\pi^+ V\to \pi e^+ e^-$ may come from the resonance.
Thus, we require (\ref{Ktopi}) be less than  $3\times 10^{-8}$, and 
 arrive at the constraint on mass versus coupling plotted in Figure 2. As one can see,
 the constraint becomes stronger than $(g-2)_\mu$ for $m_V$ around 300 MeV. 
 We also include a sensitivity line, up to which the model can be probed if $\Delta {\rm Br}_{\rm res}\sim 6\times 
10^{-9}$ can be achieved in the dedicated analysis of lepton spectra.

 Among fully leptonic decays of type B (\ref{Kproc}), we choose the one that is technically the 
 simplest, $K^+\to e^+\nu V$ and is analogous to $K^+\to e^+\nu \gamma$. Because of the 
electron chirality suppression of $K^+\to e^+\nu$,
 the $\gamma$ or $V$ bosons have to be radiated by the structure-dependent vertex, {\em i.e.} 
 not by the initial kaon or final positron line. This simplifies our task, given 
that radiative leptonic decays of pseudoscalars are reasonably well understood. In order 
 to estimate the branching ratios in the limit of small $m_V$, $m_V \la 200$ MeV,
 we simply multiply the SM rate by the mixing parameter $\kappa^2$,
\be
\label{eenu}
{\rm Br}_{K^+\to l^+\nu V} \simeq \kappa^2  {\rm Br}_{K^+\to l^+\nu \gamma} = 1.5\times 10^{-5}\times \kappa^2 ,
\ee
while for heavier $m_V$ one would need to perform a separate calculation to 
include contributions from other $q^2$-proportional form factors and account for 
the phase space suppression. We notice, however, that from (\ref{eenu}) one can immediately conclude that 
for $\kappa^2 < 10^{-3}$ the branching ratios fall below $10^{-8}$ level, which is comparable 
to the experimental errors on the branching ratios of the SM processes $K^+\to l^+\nu l^+l^-$.
We conclude that constraints on mixing provided by processes of type B (\ref{Kproc}) 
are  subdominant to muon $g-2$ constraints, and therefore there is no 
pressing need in refining estimate (\ref{eenu}).

\subsection{ Radiative hyperon decay, HyperCP anomaly, and the hypothesis of 214 MeV boson} 

$V$-boson can also be produced in the radiative hyperon decays. Since the branching ratio 
for the SM radiative decay is very large, ${\rm Br}_{\Sigma^+\to p\gamma} = 1.2 \times 10^{-3}$,
one can expect an enhanced rate for the production of $V$ vector boson. The dedicated search of 
$\Sigma^+\to p\mu^+\mu^-$ process has produced some unexpected results \cite{HyperCP}: three observed events
were consistent with the expectation for the SM rate, but their spectrum, 
all three clustered around $m_{\mu\mu}=214.3\pm$0.5 MeV,
is extremely puzzling. The HyperCP collaboration estimates that the probability of this happening within
SM is less than 1\%, while the hypothesis of the two-body decay, $\Sigma^+\to pX$, with $M_X\simeq 214.3$MeV
followed by subsequent immediate decay $X\to \mu^+\mu^-$ can account for the anomalous 
energy distribution, and allows to extract the $X$-mediated branching ratio. 
This defines the "HyperCP hypothesis" that consists of
\be
\label{whatyouneed}
{\rm HyperCP~ hypothesis:}~~~
m_X=214.3~{\rm MeV};~~~{\rm Br}^X_{\Sigma^+\to p\mu^+\mu^-} = 3.1^{+2.4}_{-1.9}({\rm stat}) \pm 1.5({\rm syst}).
\ee
This hypothesis generated the whole line of theoretical investigations 
that revisited SM calculations of semi-leptonic $\Sigma^+$ decays \cite{Valencia1}, 
made a general analysis of possible New Physics 
contributions \cite{Valencia2}, and invested some (semi-convincing) model-building 
efforts in an attempt to "find" $X$-particles within more defined models of New Physics \cite{mogila}.

Ref. \cite{Valencia2} analyzes possible couplings of "HyperCP"-boson and concludes 
that bosons with vector or scalar couplings to $s-d$ flavor-changing 
currents are not allowed as an explanation of \cite{HyperCP}, while pseudoscalars and axial-vectors
are possible. The negative conclusion with respect to vector-coupled $X$ 
comes from the analysis of $K^+ \rightarrow \pi^+ l^+ l^-$ decays. 
Very naively, this analysis would then preclude $V$ boson of secluded \uop\ that has only 
vector couplings to serve as a candidate for $X$. We find that this conclusion 
does not apply to our model, because the 
results of Ref. \cite{Valencia2} rest on the assumption 
that New Physics in flavor sector is dominated by the {\em short-distance}
contributions, such as $X_\mu \bar s \gamma_\mu d$ and alike. This is not the case 
for the secluded \uop\ model, where New Physics in form of $V$-bosons 
couples entirely through the mixing with photons and thus through the 
{\em long-distance} effects. It is widely known that the long-distance contributions 
dominate the short-distance ones in the radiative decays of $\Sigma$ by as much as three 
orders of magnitude. In what follows, we  investigate whether the putative "HyperCP" 
boson (\ref{whatyouneed}) can be identified with $V$, find the acceptable mixing $\kappa$ 
that provides required rate for $\Sigma^+\to p V \to p \mu^+\mu^-$ process,
and compare this prediction with other constraints. 

We start from the standard parametrization of the $\Sigma^+\to p \gamma$ decay form factors 
$a$, $b$ $c$ and $d$ featured in the matrix element for $\Sigma\to p$ electromagnetic transition:
\be
{\cal M}_{\rm \Sigma\to p\gamma^*} = 
eG_F \bar p [i\sigma_{\mu\nu}q_\mu (a+b\gamma_5) +(q^2\gamma^\nu -q^\nu \slash q)(c+d\gamma_5)]  \Sigma .
\label{Seff}
\ee
Here we follow the convention of Ref. \cite{Valencia1}.
For the emission of a real photon only $a$ and $b$ form factors at $q^2=0$ are relevant,
\be
\Gamma_{\Sigma^+\to p \gamma} = \fr{G_F^2e^2}{\pi}\left(|a(0)|^2 + |b(0)|^2 \right) E_\gamma^3.
\ee
This rate,  upon normalization  on relevant branching ratio of $1.2\times 10^{-3}$,
gives the following inference about the size of the form-factors \cite{Valencia1}:
\be
|a(0)|^2 + |b(0)|^2 = (15\pm 0.3~{\rm MeV})^2;~~~ {\rm Re}(a(0)b^*(0)) = (-85\pm 9.6) ~{\rm MeV}^2.
\ee
The second relation comes from the measurement of the parity-violating interference in $\Sigma$ decay. 
Unfortunately, only the cursory information can be gathered about eight form factors, counting 
real and imaginary parts, as functions of $q^2$. In the assumption of $q^2$-independence,
conjectured from the mild $q^2$-dependence reconstructed for imaginary parts of the form factors 
with the use of chiral perturbation/heavy baryon theory \cite{Valencia1}, the prediction for the 
SM branching ratio is:
\ba
\nonumber
10^8 \times {\rm Br}^{\rm SM}_{\Sigma^+\to p \mu^+\mu^-} \sim 4.5\times 
\fr{|a|^2+|b|^2}{(15~{\rm MeV})^2} -3.6\times \fr{|a|^2-|b|^2}{(15~{\rm MeV})^2}
\;\;\;\;\;\;\;\;\;\;\;\;\;\;
\\
\label{SMSi}
+1.1\times \left(\fr{|c|}{0.1}\right)^2 +0.18\times \left(\fr{|d|}{0.01}\right)^2
+0.45\times \fr{{\rm Re}(ac^*)}{1.5~{\rm MeV}} - 2.4\times \fr{{\rm Re}(bd^*)}{0.15~{\rm MeV}}
\;\;\;\;\;\;\;\;\;\;\;\;\;\;\\
=4.5\left[  |A|^2 + |B|^2 - 0.8(|A|^2 - |B|^2) +0.33|C|^2 +0.04|D|^2
+0.1{\rm Re}(AC^*) -0.54{\rm Re}(BD^*)
 \right].
\nonumber
\ea
We took a liberty of normalizing form factors $a,~b,~c,~d$ on their typical values inferred in 
\cite{Valencia1}, and introduced dimensionless and (presumably) $O(1)$ 
values $A,~B,~C,~D$. It is important to keep in mind that even if one adopts the 
chiral perturbation inspired determinations of real and imaginary parts of  $a$ and $b$, 
there exists a residual ambiguity in the choice of signs, that results in (\ref{SMSi}) 
covering  an entire range $10^{-9} \la  {\rm Br}^{\rm SM}_{\Sigma^+\to p \mu^+\mu^-} \la 10^{-8} $ \cite{Valencia1}, 
and perhaps even wider if one adopts more conservative assumptions about theoretical errors in extracting the 
form factors. One cannot help noticing numerous possibilities for constructive or 
destructive interference between different terms, which are beyond theoretical control
at the current stage of our understanding of strong dynamics. 

The calculation of the $\Sigma \to pV$ decay in terms of form factors in (\ref{Seff}) is similar
if not simpler, and we quote 
our result directly for the case of $m_V=214.3$ MeV, 
\ba
\label{BrHP}
{\rm Br}_{\Sigma^+\to p V} = 5.2\times 10^{-4} \kappa^2 \times N_{ABCD},
\ea
where $N_{ABCD}$ stands for the following combination  taken at $q^2 = (214.3~{\rm MeV})^2$:
\be
\label{NABCD}
N_{ABCD} = |A|^2 + |B|^2 - 0.66(|A|^2 - |B|^2) +0.35|C|^2 +0.03|D|^2
+0.14{\rm Re}(AC^*) -0.44{\rm Re}(BD^*)
\ee
One can easily see that if the rate is dominated by 
$|A|^2 + |B|^2\simeq 1$ and the rest of the contributions is small, 
the branching ratio to $V$ bosons is about $0.4\kappa^2$ of the 
$\Sigma\to p\gamma$ branching, where the factor of $0.4$ comes from the phase space suppression. 
Variation of coefficients in (\ref{NABCD}) suggests a $\sim(10^{-4}-10^{-3})\times \kappa^2$ 
branching ratio range for $\Sigma\to pV$ decay, and in order to get to the 
final estimate of $V$-mediated $\Sigma^+\to p\mu^+\mu^-$ decay, we must multiply the 
rate by ${\rm Br}_{V\to \mu^+\mu^-}\simeq 0.2$. Therefore,
we arrive at the following estimate for the HyperCP decay rate mediated by $V$ bosons,
\ba
\label{result}
{\rm Br}^V_{\Sigma^+\to p \mu^+\mu^-} = {\rm Br}_{V\to \mu^+\mu^-}\times {\rm Br}_{\Sigma^+\to p V}=
1. \times 10^{-4} \kappa^2 \times N_{ABCD}.
\ea
Assigning somewhat arbitrarily $N_{ABCD}^{max} \sim 3$, one deduces the following {\em minimal} 
value of $\kappa$ consistent with hypothesis (\ref{whatyouneed}):
\be 
\kappa^2 > 3\times 10^{-5}.
\label{whatyouget}
\ee
This lower bound of mixing compatible with HyperCP hypothesis is also consistent with $a^V_\mu \sim 3\times 10^{-9}$,
which is exactly in the middle of the band that "solves" $(g-2)_\mu$ discrepancy, Figure 2. It is fair to 
say, however, that the most natural values for $\kappa^2$ consistent with (\ref{whatyouneed}) 
are well above $10^{-4}$ and in the domain already excluded by $K^+\to \pi^+ \mu^+\mu^-$ and muon $g-2$. 
One can also notice that the coefficients in $N_{ABCD}$ are similar to those in the square bracket of (\ref{SMSi}),
although not exactly the same. 
This is the direct consequence of the fact that the phase space available for the SM muon decays is not 
that large, and on average muon pairs have an invariant mass not far from 214 MeV. 
Therefore, it would be natural to 
divide the two rates, (\ref{SMSi}) and (\ref{BrHP}), in an attempt to reduce the uncertainty. We are not 
pursuing this idea here, because in the end there is no  guarantee that the 
form factor dependence of (\ref{SMSi}) and (\ref{BrHP})
is the same, given the number of assumptions that was made on the way.

\section{Discussion}

We have presented some results on the phenomenology of \uop\ gauge boson in the mass 
range of few MeV to GeV, with the kinetic mixing to photon at $O(10^{-3}-10^{-2})$ level. 
We have concluded that none of the constraint that have been analyzed in this
paper can decisively rule out this possibility. 
Indeed, the opportunities for observing such $V$-boson are very "minimal" since 
it has  only electromagnetic couplings. As a result, the best constraints one can find come
from the precision QED experiments. As expected, the muon $g-2$ is the best source for limiting $\kappa-m_V$
parameter space at $m_V \sim 100$ MeV. However, because of the controversial status of the 
experiment vs theory, one cannot rule out $V$ bosons with per-mill couplings, and moreover, 
an additional contribution of $V$-muon loop with $\kappa^2\sim 10^{-5}-10^{-4}$ may actually improve the 
agreement between theory and experiment.

The radiative decays of strange particles have been extensively studied in the past and are 
another way of probing $m_V-\kappa$ parameter space. 
Assuming that $V$ decays back to leptons (as opposed to some unspecified $O({\rm MeV})$ matter 
charged under \uop), we have shown that $K^+\to \pi^+e^+e^-$ decay is already limiting 
the model better than $(g-2)_\mu$, but only for a rather narrow range of $m_V$ around 300 MeV. 
The decays of $K$-mesons to $V$ are dominated by the long-distance contributions,
which are significantly enhanced relative to short-distance pieces. 
The $\Sigma^+\to p l^+ l^-$ particles is another natural place to look for $V$ production. 
However, the status of $\Sigma^+\to p \mu^+ \mu^-$ decay is somewhat controversial. 
The HyperCP collaboration saw an unusual pattern of muon invariant mass distribution, with
all events (all=3) clustered around the same energy \cite{HyperCP}, and put forward a hypothesis 
that the decay is mediated by the two-body decay with some intermediate 
resonance with mass of  214.3 MeV. We have analyzed whether such hypothesis is 
viable within the secluded \uop\ model and found that given large uncertainties in evaluating 
long-distance contributions, such possibility is not excluded if $\kappa^2 > 3\times 10^{-5}$. 
The reason why this model avoids a "no-go" theorem of Ref. \cite{Valencia2}, that forbids
vector-coupled particles as possible explanation of HyperCP anomaly, is because the 
dominance of the long-distance effects, to which this theorem does not apply. 
Since the model is well-defined and the radiative kaon decays are well studied, 
the dedicated re-analysis of $K^+\to \pi^+e^+e^-$ data may close the window 
on the HyperCP hypothesis in this model. Before we conclude, a few final remarks are in order.

\begin{itemize} 

\item {\em Other realizations of MeV--GeV scale \uop.} As noted in the introduction, 
secluded \uop\ with kinetic mixing is natural, but not the only possible \uop\ extension 
below the GeV scale. If $B-L$ symmetry is gauged, low $m_V$ would require the \uop\ coupling 
constant to be much smaller than $\alpha_{EM}$ \cite{Fayet2}. The QED constraints considered 
in this paper can be simply rescaled to limit $\alpha'$. It is also plausible 
to find an appropriate $\alpha'$ that would fit HyperCP hypothesis. 
This model, however, has additional constraints related to {\em e.g.} 
neutrino interactions that might be far more important than those considered in this paper. 

\item {\em Prospects for refining constraints on mass-mixing parameter space.}
It is unlikely that one can achieve further breakthroughs either in theory 
or experiments studying radiative decays of strange particles. 
However, a possibility of direct search for narrow (a factor of a million more narrow than $J/\psi$!)
resonances in $e^+e^-$ machines should not be discarded as a way of limiting the parameter space 
of the model. Ultimately, only this could decisively test small values of $\kappa$ 
down to $10^{-3}$ level for models
with $m_V \la$GeV. 

\item {\em Doubly-secluded Higgs$'$ as a source of leptons.} So far the physics of Higgs$'$ boson has been ignored. 
It is worth pointing out, however, that Higgs$'$ decay properties are very sensitively dependent
on its mass relative to $m_V$. One can easily see that there are three main regimes for the Higgs$'$
decay to leptons (assuming $m_l \ll m_{h'}$ for simplicity):
\ba
\nonumber
 2m_V < m_{h'}:~~~ h'\to 2V;~~~\Gamma_{h'} \sim O(\kappa^0)\\
 \nonumber
m_V < m_{h'} < 2m_V :~~~ h'\to Vl^+l^-;~~~\Gamma_{h'} \sim O(\kappa^2)\\
 m_{h'} < m_V :~~~ h'\to l^+l^-l^+l^-;~~~\Gamma_{h'} \sim O(\kappa^4)
 \label{doublysec}
\ea
The last line of (\ref{doublysec}) corresponds to the regime when the Higgs 
can decay only via a pair of virtual $V$-bosons, each of which would have to decay electromagnetically
via the mixing with photon.
As a result, the amplitude of Higgs$'$ decay is quadratic in $\kappa$ and the width is quartic, making this 
Higgs "doubly-secluded". Being further suppressed by $\alpha^2$ and the four-particle phase space, 
the lifetime of Higgs$'$ can be much longer than the lifetime of $V$ and indeed on the 
order of the lifetimes of particles that decay due to weak interactions. It is also worth pointing
out that the production cross section of $h'$ by Higgs-strahlung mechanisms 
is only singly-secluded, $\sigma_{f\bar f \to Vh'} \sim O(\kappa^2)$. Interestingly, a scenario of
a particle with not very suppressed production rate and very small decay rate was suggested
recently by the CDF collaboration in connection with observation of "ghost muons" \cite{CDF}. At some superficial level,
the doubly-secluded Higgs with $m_{h'} > 4 m_\mu$ may fit this scenario, 
but without a dedicated analysis it is 
impossible to tell whether the production rate of $h'$ could be large enough to account for 
"extra CDF muons". On the other hand, the Higgs$'$-strahlung signature of 6 leptons in the 
final state can be used in $e^+e^-$ machines to set additional constraints on the model. 

\end{itemize}

\subsection*{Acknowledgments}

The author would like to thank Saveliy Karshenboim, Mikhail Voloshin and Adam Ritz 
for useful discussions.
Research at the Perimeter Institute
is also supported in part by the Government of Canada through NSERC and by the Province
of Ontario through MEDT.


\begin{thebibliography}{99}

\bibitem{Holdom}
  B.~Holdom,
  Phys.\ Lett.\  B {\bf 166}, 196 (1986).
  
     \bi{PRV1}
  M.~Pospelov, A.~Ritz and M.~B.~Voloshin,
  Phys.\ Lett.\  B {\bf 662}, 53 (2008)
  [arXiv:0711.4866 [hep-ph]].
  
  \bi{Fayet1} P.~Fayet,
  Nucl.\ Phys.\  B {\bf 347}, 743 (1990).
  
    
  \bibitem{Langacker}
  P.~Langacker,
  arXiv:0801.1345 [hep-ph].

  
  \bi{BF} C.~Boehm and P.~Fayet,
  Nucl.\ Phys.\  B {\bf 683}, 219 (2004)
  [arXiv:hep-ph/0305261].
  
  
 \bi{Kim} J.~H.~Huh, J.~E.~Kim, J.~C.~Park and S.~C.~Park,
  Phys.\ Rev.\  D {\bf 77}, 123503 (2008)
  [arXiv:0711.3528 [astro-ph]].
  
\bibitem{BouchF}
  C.~Bouchiat and P.~Fayet,
  Phys.\ Lett.\  B {\bf 608}, 87 (2005)
  [arXiv:hep-ph/0410260].

  
  
 \bi{Fayet2}  P.~Fayet, Phys.\ Rev.\  D {\bf 74}, 054034 (2006)
  [arXiv:hep-ph/0607318]; P.~Fayet,
  Phys.\ Rev.\  D {\bf 75}, 115017 (2007)
  [arXiv:hep-ph/0702176].
 

\bi{recent_stuff}
 N.~Arkani-Hamed, D.~P.~Finkbeiner, T.~Slatyer and N.~Weiner,
  arXiv:0810.0713 [hep-ph]; 
 N.~Arkani-Hamed and N.~Weiner,
  arXiv:0810.0714; 
M.~Pospelov and A.~Ritz,
  arXiv:0810.1502 [hep-ph].

\bi{Pamela}  O.~Adriani {\it et al.},
  arXiv:0810.4995 [astro-ph].


\bi{lowV} A.~E.~Nelson and J.~Walsh,
  Phys.\ Rev.\  D {\bf 77}, 033001 (2008)
  [arXiv:0711.1363 [hep-ph]]; 
  J.~Jaeckel, J.~Redondo and A.~Ringwald,
  Phys.\ Rev.\ Lett.\  {\bf 101}, 131801 (2008)
  [arXiv:0804.4157 [astro-ph]]; 
   M.~Pospelov, A.~Ritz and M.~B.~Voloshin,
  arXiv:0807.3279 [hep-ph]; M.~Ahlers, J.~Jaeckel, J.~Redondo and A.~Ringwald,
  Phys.\ Rev.\  D {\bf 78}, 075005 (2008)
  [arXiv:0807.4143 [hep-ph]].
  
  \bi{Harvard} B. Odom, D. Hanneke, B. D'Urso, and G. Gabrielse,  
Phys.\ Rev.\ Lett.\  {\bf 97}, 030801 (2006).

\bi{GK} G. Gabrielse, D. Hanneke, T. Kinoshita, M. Nio, and B. Odom,
Phys.\ Rev.\ Lett.\  {\bf 97}, 030802 (2006); Erratum: Phys.\ Rev.\ Lett.\  {\bf 99}, 039902 (2007).

\bi{CsRb} P. Clad\'e {\em et al.}, Phys.\ Rev.\ Lett.\  {\bf 96}, 033001 (2006);
V. Gerginov {em et al.}, Phys.\ Rev.\ A  {\bf 73}, 032504 (2006).
 
  \bibitem{RS}
  P.~D.~Serpico and G.~G.~Raffelt,
  Phys.\ Rev.\  D {\bf 70}, 043526 (2004)
  [arXiv:astro-ph/0403417].
  
  \bibitem{PMS} M.~Passera, W.~J.~Marciano and A.~Sirlin,
  arXiv:0809.4062 [hep-ph].

  \bibitem{gmu} G.~W.~Bennett {\it et al.}  [Muon G-2 Collaboration],
  Phys.\ Rev.\  D {\bf 73}, 072003 (2006)
  [arXiv:hep-ex/0602035].
  
  \bi{tau} M.~Davier, S.~Eidelman, A.~Hocker and Z.~Zhang,
  Eur.\ Phys.\ J.\  C {\bf 27}, 497 (2003)
  [arXiv:hep-ph/0208177].

  
  \bibitem{DavierNsk} M. Davier, talk at TAU'2008 meeting, Novosibirsk, September 2008;
  http://tau08.inp.nsk.su/prog.php
  
  \bibitem{Sav} S.~G.~Karshenboim,
  Phys.\ Rept.\  {\bf 422}, 1 (2005)
  [arXiv:hep-ph/0509010].
  
  \bi{Bird} C.~Bird, P.~Jackson, R.~V.~Kowalewski and M.~Pospelov,
  Phys.\ Rev.\ Lett.\  {\bf 93}, 201803 (2004)
  [arXiv:hep-ph/0401195]; C.~Bird, R.~V.~Kowalewski and M.~Pospelov,
  Mod.\ Phys.\ Lett.\  A {\bf 21}, 457 (2006)
  [arXiv:hep-ph/0601090].

  
  \bibitem{dAmbrosio} 
  G.~D'Ambrosio, G.~Ecker, G.~Isidori and J.~Portoles,
  JHEP {\bf 9808}, 004 (1998)
  [arXiv:hep-ph/9808289].

\bibitem{Appel} R.~Appel {\it et al.}  [E865 Collaboration],
  Phys.\ Rev.\ Lett.\  {\bf 83}, 4482 (1999)
  [arXiv:hep-ex/9907045].
  
  \bibitem{Park}H.~K.~Park {\it et al.}  [HyperCP Collaboration],
  Phys.\ Rev.\ Lett.\  {\bf 88}, 111801 (2002)
  [arXiv:hep-ex/0110033].



\bibitem{invisible} V.~V.~Anisimovsky {\it et al.}  [E949 Collaboration],
  Phys.\ Rev.\ Lett.\  {\bf 93}, 031801 (2004)
  [arXiv:hep-ex/0403036].
  
  \bi{HyperCP} H.~Park {\it et al.}  [HyperCP Collaboration],
  Phys.\ Rev.\ Lett.\  {\bf 94}, 021801 (2005)
  [arXiv:hep-ex/0501014].

\bi{Valencia1} X.~G.~He, J.~Tandean and G.~Valencia,
  Phys.\ Rev.\  D {\bf 72}, 074003 (2005)
  [arXiv:hep-ph/0506067].

\bi{Valencia2} 
  X.~G.~He, J.~Tandean and G.~Valencia,
  Phys.\ Lett.\  B {\bf 631}, 100 (2005)
  [arXiv:hep-ph/0509041].

\bi{mogila} N.~G.~Deshpande, G.~Eilam and J.~Jiang,
  Phys.\ Lett.\  B {\bf 632}, 212 (2006)
  [arXiv:hep-ph/0509081]; D.~S.~Gorbunov and V.~A.~Rubakov,
  Phys.\ Rev.\  D {\bf 73}, 035002 (2006)
  [arXiv:hep-ph/0509147]; C.~Q.~Geng and Y.~K.~Hsiao,
  Phys.\ Lett.\  B {\bf 632}, 215 (2006)
  [arXiv:hep-ph/0509175];  X.~G.~He, J.~Tandean and G.~Valencia,
  Phys.\ Rev.\ Lett.\  {\bf 98}, 081802 (2007)
  [arXiv:hep-ph/0610362]; C.~H.~Chen, C.~Q.~Geng and C.~W.~Kao,
  Phys.\ Lett.\  B {\bf 663}, 400 (2008)
  [arXiv:0708.0937 [hep-ph]]; 
   X.~G.~He, J.~Tandean and G.~Valencia,
  JHEP {\bf 0806}, 002 (2008)
  [arXiv:0803.4330 [hep-ph]].

\bi{CDF} T.~Aaltonen {\it et al.},
  arXiv:0810.5357 [hep-ex].

  
  \end{thebibliography}
\end{document}